\documentclass[prb,onecolumn,amsmath,amssymb, superscriptaddress, nofootinbib]{revtex4}
\usepackage[pdftex]{graphicx}
 \usepackage{amsmath}
\usepackage[T1]{fontenc}
\usepackage{amssymb}
\usepackage{amsfonts}
\usepackage{bm}
 \usepackage{amsmath} 
  \usepackage{flushend}
\usepackage{color}
\usepackage{amsfonts} 
\usepackage{amssymb, mathrsfs}
\usepackage{braket}
\usepackage{graphicx} 
\usepackage{subfigure}
\usepackage{bm} 
\usepackage{bbm}
\def\beq{\begin{equation}}
\def\eeq{\end{equation}}
\def\bsp{\begin{split}}
\def\esp{\end{split}}
\def\bea{\begin{eqnarray}}
\def\eea{\end{eqnarray}}
\def\ba{\begin{array}}
\def\ea{\end{array}}

\def\dg{\dagger}

\def\lb{\left(}
\def\rb{\right)}

\def\l.{\left.}
\def\r.{\right.}

\def\ra{\rangle}
\def\la{\langle}

\def\bo{{\vec k}}

\begin{document}

\date{\today}
\title{\Large Magnonic Floquet Quantum Spin Hall Insulator in Bilayer Collinear Antiferromagnets}
\email{sowerre@perimeterinstitute.ca}
\author{S. A. Owerre}
\affiliation{Perimeter Institute for Theoretical Physics, 31 Caroline St. N., Waterloo, Ontario N2L 2Y5, Canada.}

\maketitle

\noindent\textbf{
We study irradiated two-dimensional insulating bilayer honeycomb ferromagnets and antiferromagnets coupled antiferromagnetically with a zero net magnetization.  The former is realized in the recently synthesized bilayer honeycomb chromium triiodide  CrI$_{\bf 3}$.  In both systems, we show that circularly-polarized electric field breaks time-reversal symmetry and induces a dynamical  Dzyaloshinskii-Moriya  interaction in each  honeycomb layer.  However, the resulting bilayer antiferromagnetic system still preserves a combination of  time-reversal and space-inversion ($\boldsymbol{\mathcal{PT}}$)  symmetry.  We show that the magnon topology of the bilayer antiferromagnetic system is characterized by a $\pmb{\mathbb{Z}_2}$ Floquet topological invariant. Therefore, the system realizes a  magnonic Floquet quantum spin Hall insulator with spin filtered magnon edge states. This leads to  a non-vanishing Floquet magnon spin Nernst effect, whereas the Floquet magnon thermal Hall effect vanishes due to $\boldsymbol{\mathcal{PT}}$ symmetry.  We study the rich $\pmb{\mathbb{Z}_2}$ Floquet topological magnon phase diagram of the system as a function of the light amplitudes and polarizations.  We further discuss the great impact of the results on future experimental realizations.}

\vspace{10px}

In recent years, periodically driven solid-state materials have emerged as an alternative avenue to extend the search for topological quantum materials \cite{foot3,foot4,foot5,gru,fot,fot1,jot,fla,we1,we2,we3,we4,we5,we6, gol,buk,eck1,ste,ple,ew, dik, lin, du,du1, delp, eza, zhai, saha,roy,roy1}. This mechanism involves the exposure of a topologically trivial quantum material to a time-periodic electric field.  In this system, time-reversal symmetry of the Bloch bands  is broken by circularly-polarized electric field  by modifying  the  intrinsic properties of the material via light-matter interactions.  This results in a Floquet Chern insulator such as in irradiated graphene \cite{foot3,foot4}. The non-equilibrium topological systems are believed to give interesting properties that are not possible in the equilibrium systems.

In insulating magnets, the quantum theory of magnons dictates that magnons carry a spin magnetic dipole moment and an intrinsic spin of $1$, which can be used for dissipationless information processing in the emerging field of magnon spintronics \cite{benja,magn}. This implies that magnons can accumulate the Aharonov-Casher phase \cite{aha,ahaz,ahat,mei,mei1,tak4} when exposed to a time-independent spatially-varying electric field  resulting in magnonic Landau levels \cite{mei} and chiral anomaly in Weyl magnons \cite{xr, xr1}. Remarkably, the magnon accumulated Aharonov-Casher phase has a strikingly different physics when the  electric field is time-dependent and periodic as in electronic systems. In this case, the resulting irradiated insulating magnets can be investigated using the Floquet theory in a similar manner to irradiated metallic electronic systems. Unlike electronic systems, the magnetic Floquet physics can reshape the underlying spin Hamiltonian to stabilize magnetic phases  and provides a promising avenue for  inducing and tuning Floquet topological spin excitations \cite{owe0,owe2,kar}, with a direct implication  of generating and manipulating ultrafast spin current using terahertz ({\rm THz})  radiation \cite{ment, tak4a,walo}.  In this respect, the concept of magnonic Floquet Chern insulator has emerged \cite{owe0,owe2,kar}, where circularly-polarized light induces a dynamical Dzyaloshinskii-Moriya (DM) interaction \cite{sol,dm,dm2}  in a single-layer two-dimensional (2D) insulating honeycomb ferromagnet. This approach has also been generalized to engineer  Floquet  Weyl magnons \cite{owe1} in three-dimensional (3D) insulating honeycomb ferromagnets. Similar to electronic Floquet system, time-reversal symmetry is broken by circularly-polarized electric field and the Floquet topological system  is characterized by the first Chern number. Therefore, the topological aspects of electronic and magnonic Floquet  systems are essentially the same and they originate from the same oscillating time-periodic electric field. To make this similarity obvious, we note that in the magnonic Floquet topological systems,  the intensity of light is characterized by the dimensionless quantity  
 \bea 
 \mathcal E_i = \frac{g\mu_B E_i a}{\hbar c^2},
 \label{eq1}
 \eea 
 where $E_i$ ($i= x,y$) are the amplitudes of the electric  field, $g$ is the Land\'e g-factor, $\mu_B$ is the Bohr magneton,  $a$ is the lattice constant, $\hbar$ and $c$ are the reduced Plank's constant and the speed of light respectively.  The dimensionless quantity in Eq.~\eqref{eq1} should be compared to that of electronic Floquet topological systems \cite{foot3, foot5,foot4}
 \bea 
 \mathcal E_i = \frac{e E_i a}{\hbar \omega},
 \eea 
 where $e$ is the electron charge and $\omega$ is the angular frequency of light.  Thus, for the irradiated magnetic insulators we can identify the spin magnetic dipole moment carried by magnon as 
 \bea 
 g\mu_B = \frac{ec^2}{\omega} =\frac{ec\lambda}{2\pi}.
 \eea 
Therefore, we can see that for a typical light wavelength $\lambda$ of order $10^{-8}$m, the spin magnetic dipole moment $ g\mu_B$ carried by magnon in the irradiated magnetic insulators is comparable to the electron charge $e$. This shows the similarity between the electronic and the magnonic Floquet topological systems.

Recently, the $\mathbb{Z}_2$ characterization of topological magnon bands in the equilibrium time-independent insulating antiferromagnets  has garnered considerable attention \cite{z1,z2,z3,z4,z5,z6}.  In particular, for the 2D insulating bilayer honeycomb antiferromagnets with a DM interaction \cite{sol,dm,dm2} the system preserves time-reversal and space-inversion $(\mathcal{PT})$ symmetry and realizes the magnonic  analog of  $\mathbb{Z}_2$ topological insulator \cite{qs,qs1,qs2}. Unfortunately, most 2D insulating honeycomb antiferromagnets do not have the unique form of the required DM interaction \cite{sol}. In fact, the absence of this unique DM interaction in most insulating honeycomb antiferromagnets has prevented a discernible experimental observation of the magnon spin Nernst voltage in MnPS$_3$ \cite{shi}. One possible mechanism to induce the unique form of the required DM interaction  in 2D insulating honeycomb antiferromagnets is through photo-irradiation with a circularly-polarized electric field \cite{owe0}.

In this report, we propose a $\mathbb{Z}_2$ magnonic Floquet quantum spin Hall insulator in  bilayer collinear antiferromagnets with $\mathcal{PT}$ symmetry.  Specifically,  we study irradiated 2D insulating bilayer honeycomb ferromagnets and antiferromagnets coupled antiferromagnetically with a zero net magnetization, where the former is realized in bilayer CrI$_3$ \cite{Huang, Huang1, Huang2}. Our theoretical formalism is based on the Floquet theory, spin-wave theory, and quantum field theory.   In both honeycomb bilayer systems, we show that circularly-polarized electric field induces a dynamical  DM   interaction in each honeycomb layer, but the bilayer antiferromagnetic system preserves  $\mathcal{PT}$ symmetry, hence the resulting magnon topology is  characterized by a $\mathbb{Z}_2$ Floquet topological invariant quantity. We obtain the  $\mathbb{Z}_2$ Floquet topological magnon phase diagram and identify the regimes where the Floquet magnon spin Nernst coefficient changes sign.  We also show that both systems exhibit Floquet spin-filtered magnon edge states, where Floquet magnon with opposite spin propagates in opposite directions. Our results provide a powerful mechanism for manipulating the intrinsic properties of 2D insulating  honeycomb antiferromagnetic materials such as  bilayer CrI$_3$, and could pave the way for studying new interesting features in 2D insulating antiferromagnets such as  photo-magnonics \cite{benj}, magnon spintronics \cite{magn, benja},  and ultrafast optical control of magnetic spin currents \cite{ment, tak4, tak4a,walo}.

 \vspace{10px}
\noindent \textbf{\large Results}
 
\begin{figure}
\centering
\includegraphics[width=1\linewidth]{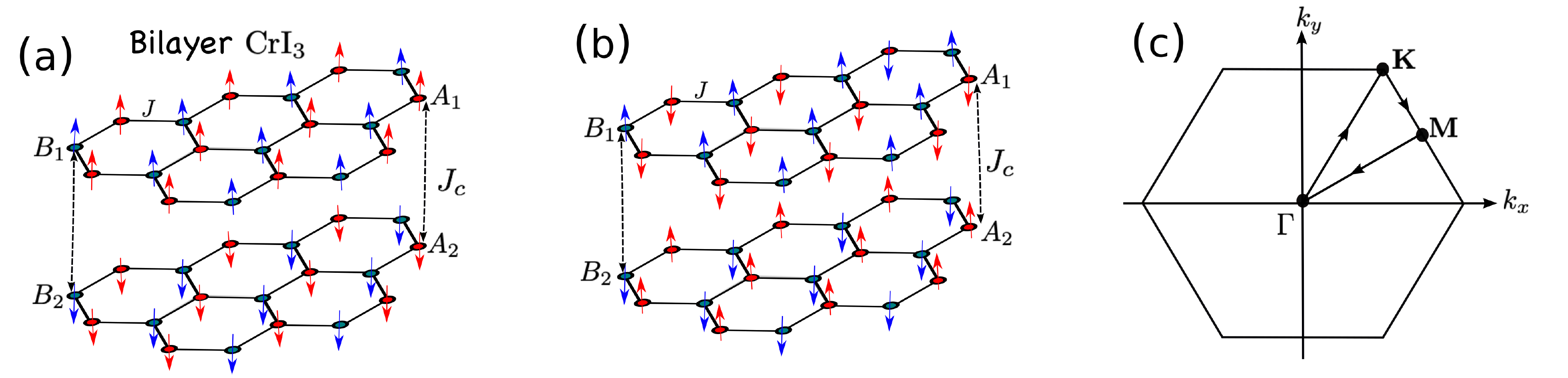}
\caption{Color online. (a) Case $(i)$ -- bilayer honeycomb-lattice ferromagnets coupled antiferromagnetically  as realized in bilayer CrI$_3$. (b) Case $(ii)$ -- bilayer honeycomb-lattice antiferromagnets coupled antiferromagnetically.  (c) The Brillouin zone (BZ) of the honeycomb lattice with high-symmetry paths.}
\label{lattice}
\end{figure}

\noindent\textbf{Bilayer Heisenberg spin model.}~~ We consider the Heisenberg spin model for 2D insulating bilayer honeycomb ferromagnets and antiferromagnets  coupled antiferromagnetically. The Hamiltonian is governed by  

 \begin{align}
\mathcal H&=J\sum_{ \la ij\ra,\ell }{\vec S}_{i,\ell}\cdot{\vec S}_{j,\ell}+ J_c\sum_{i}{\vec S}_{i}^{T}\cdot{\vec S}_{i}^{B},
\label{spinh}
\end{align}
where ${\vec S}_{i}$ is the spin vector at site  $i$ and $\ell$ labels the top ($T$) and bottom ($B$) layers.  We will consider two different cases: $(i)$ $J<0, J_c >0$ (see Fig.~\ref{lattice}(a)).  $(ii)$ $J>0, J_c >0$ (see Fig.~\ref{lattice}(b)).  The intralyer coupling is ferromagnetic in case $(i)$ and antiferromagnetic in case $(ii)$. In both cases  the net magnetization vanishes. We note that case $(i)$ is manifested in the  bilayer honeycomb magnet CrI$_3$ \cite{Huang, Huang1, Huang2}. There are four sublattices in the unit cell denoted  by $A_1, B_1, A_2, B_2$. In both cases the interlayer exchange couples sites on the sublattices $A_1 \leftrightarrow A_2$ and $B_1 \leftrightarrow  B_2$.

 \vspace{10px}
\noindent\textbf{Bosonic Bogoliubov-de Gennes model.}~~We will focus on the low-temperature regime, when  the magnetic excitations of the spin Hamiltonian   in Eq.~\eqref{spinh}  can be described  by  the Holstein Primakoff transformation \cite{hp} (see Methods). The bosonic Hamiltonian in momentum space is given by
  
\begin{align}
\mathcal H= \frac{1}{2}\sum_{\vec k}\big(u^\dg(\vec k), u(-\vec k)\big)\cdot
\mathcal H(\vec k)\cdot
{u(\vec k)\choose u^\dg(-\vec k)}.
\end{align}
The bosonic Bogoliubov-de Gennes (BdG) Hamiltonian is given by
\begin{align}
\mathcal H(\vec k)= 
\begin{pmatrix}
\mathcal H_+(\vec k) &0\\
0& \mathcal H_-(\vec k)
\end{pmatrix},
\label{bosonicH}
\end{align}
with $\mathcal{H}_{-}(\bo)=\mathcal{H}_{+}^*(-\bo)$. Each block Hamiltonian is a $4\times 4$ Hermitian matrix representing the $S_z=+1$ and the $S_z=-1$ spin sectors.  Therefore, $\mathcal{H}(\bo)$ is invariant under $\mathcal{PT}$ symmetry given by $\mathcal{PT}=\sigma_x\otimes\sigma_0 \mathcal{K}$, where $\mathcal{K}$ is complex conjugation, $\sigma_i$ are Pauli matrices with an identity  $\sigma_0$. 
The  $S_z=+1$ sector Hamiltonian in case $(i)$ ($J<0, J_c >0$) is given by
\begin{align}
\mathcal H_+^{(i)}(\bo)&= 
\begin{pmatrix}
|v_0| &-|v_J|f^*(\vec k)& v_{J_c}&0\\
-|v_J|f(\vec k) &|v_0|& 0&v_{J_c}\\
v_{J_c} &0& |v_0| &-|v_J|f(\vec k)\\
0 & v_{J_c}&-|v_J|f^*(\vec k)&|v_0|
\end{pmatrix},
\label{honn}
\end{align}
 with   $u^\dg(\vec k) = \big(a_{\vec k A_1}^\dg,a_{\vec k B1}^\dg, a_{-\vec k A_2},a_{-\vec k B2}\big)$.  The  $S_z=+1$ sector Hamiltonian in case $(ii)$ ($J>0, J_c >0$) is given by
\begin{align}
\mathcal H_+^{(ii)}(\bo)&= 
\begin{pmatrix}
v_0 &0& v_Jf^*(\vec k)&v_{J_c}\\
0 &v_0& v_{J_c}&v_Jf(\vec k)\\
v_Jf(\vec k) &v_{J_c}& v_0&0\\
v_{J_c} & v_Jf^*(\vec k)&0&v_0
\end{pmatrix},
\label{honn1}
\end{align}
 with   $u^\dg(\vec k) = \big(a_{\vec k A_1}^\dg,a_{\vec k B2}^\dg, a_{-\vec k B_1},a_{-\vec k A2}\big)$.

\begin{figure}
\centering
\includegraphics[width=.9\linewidth]{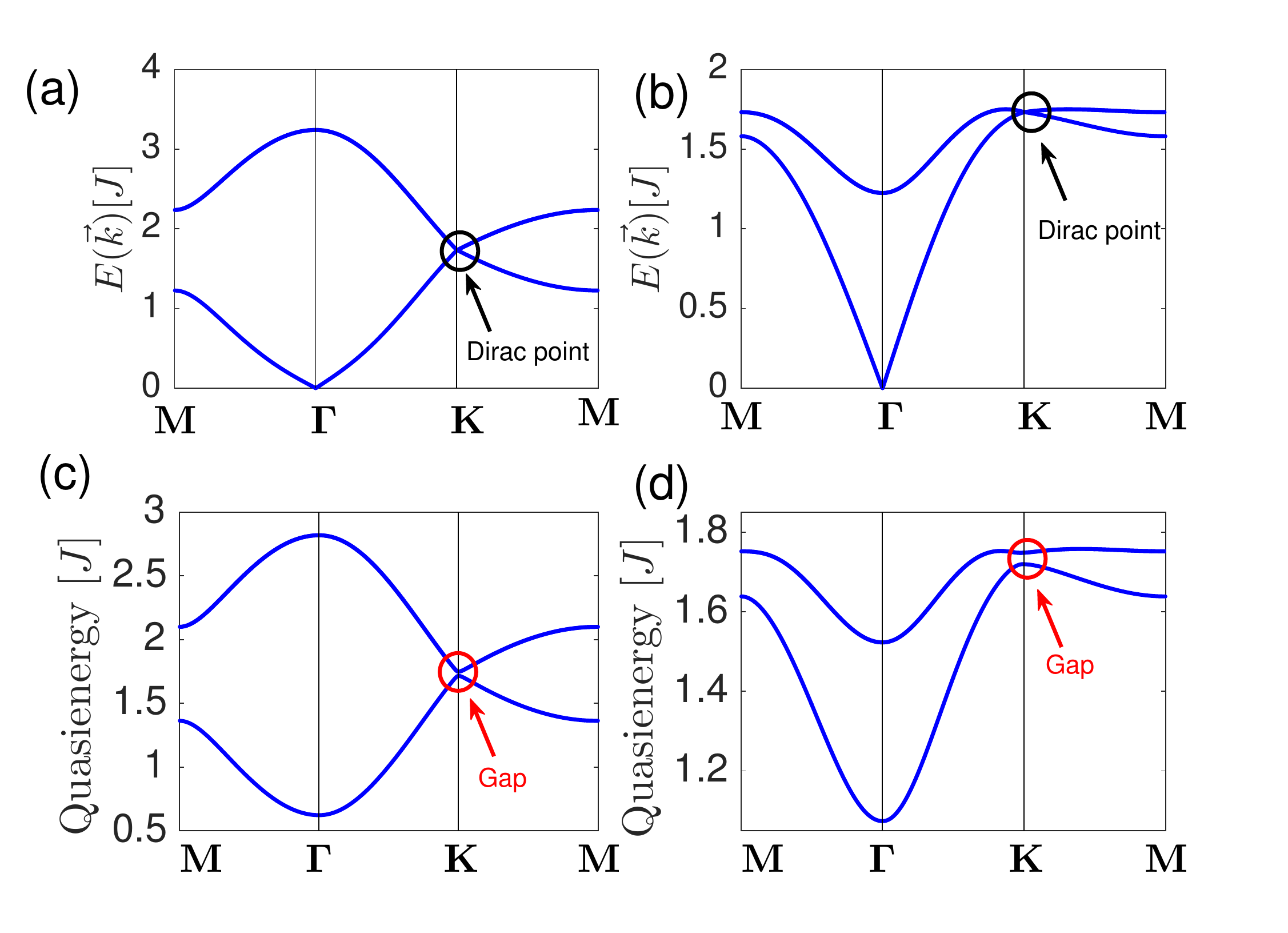}
\caption{Color online. Top panel.  Equilibrium doubly-degenerate antiferromagnetic Dirac magnon  bands for (a)  case $(i)$  and (b) case $(ii)$  with $J_c/J = 0.5$ and $S=1/2$. The Dirac points are  indicated by circles. Bottom panel. Floquet doubly-degenerate topological magnon bands for (c) case $(i)$ and  (d) case $(ii)$. The parameters are $J_c/J = 0.5$, $S=1/2$, $\mathcal{E}_x=\mathcal{E}_y = 1$, $\phi = \pi/2$, and $\hbar\omega/J = 10$. The red circle points indicate the massive Floquet Dirac magnon.}
\label{quasienergy}
\end{figure}

%
Here $v_0=3v_J + v_{J_c}$,  $v_J = JS$, $v_{J_c} = J_cS$ and $f(\vec k)=\sum_{\ell}e^{ik_\ell}$ with $k_\ell= \vec k\cdot \vec {a}_\ell$. The primitive lattice vectors   are   $\vec a_{1}=a\sqrt{3}\hat x$, $\vec a_{2}=a(\sqrt{3}\hat x/2 + 3\hat y/2)$, $\vec a_{3} =0$. The nearest-neighbour vectors are $\vec{\delta}_{1,2}=a(\mp\sqrt{3}\hat x/2 + \hat y/2)$, $\vec{\delta_3}=-a\hat y$.   The Hamiltonian can be diagonalized by paraunitary operators.  
The magnon bands are doubly-degenerate due to $\mathcal{PT}$ symmetry and they are given  by 
\begin{align}
E_{\sigma\eta}^{(i)}(\vec k) = \sqrt{\big(3v_J + |v_{J_c}| +\eta v_{J}|f(\vec k)|\big)^2 - |v_{J_c}|^2},
\end{align}
\begin{align}
E_{\sigma\eta}^{(ii)}(\vec k) = \sqrt{\big(3v_J + v_{J_c}\big)^2  -\big(v_{J_c} + \eta v_{J}|f(\vec k)|\big)^2},
\end{align}
where $\sigma = \pm $ for the layers and $\eta = \pm $ for the sublattices. The magnon energy bands are depicted in Fig.~\eqref{quasienergy}(a) and  Fig.~\eqref{quasienergy}(b) respectively.  In both cases the linear Goldstone mode at the ${\bf \Gamma}$-point signifies antiferromagnetic order. The doubly-degenerate antiferromagnetic Dirac magnons occur at the ${\bf K}$-point in both cases. 

 \vspace{10px}
 
\noindent\textbf{Irradiated bilayer antiferromagnetic insulator.}~~  
We will now  present the analysis of irradiated 2D insulating antiferromagnets.  Let us consider the effects of an oscillating  electric field ${\vec E}(\tau)$, irradiated perpendicular to the 2D insulating antiferromagnets.  The consequence of irradiated insulating antiferromagnets is that hopping magnon with  spin magnetic dipole moment $g\mu_B\hat z$ will accumulate  the time-dependent version of the Aharanov-Casher phase \cite{aha} given by (see Methods)
\begin{align}
\Phi_{ij,\ell}^s(\tau) = \hat{s}\mu_m\int_{\vec r_{i}}^{\vec r_{j}} {\vec \Xi}(\tau)\cdot d{\vec \ell},
\label{acp1}
\end{align}
where $\hat{s} = \rm{diag}(I,-I)$,  $\mu_m = g\mu_B/\hbar c^2$, and ${\vec r}_i$ is the spin position at site $i$. We have used the notation ${\vec \Xi}(\tau)={\vec E}(\tau) \times\hat z$ for brevity and ${\vec E}(\tau)=-\partial_\tau \vec{A}(\tau)$, where $\vec{A}(\tau)$ is the time-dependent vector potential given by 
 \begin{align}
 \vec A(\tau) = \big[-A_x\cos(\omega\tau +\phi),A_y\cos(\omega\tau), 0\big],
 \end{align}
where $A_{i}= E_{i}/\omega$ ($i = x,y$) are the strength of the time-dependent  vector potential and $\phi$ is the phase difference. For circularly-polarized electric field $\phi = \pi/2$ and for linearly-polarized electric field $\phi= 0$ or $\pi$.  The corresponding time-dependent oscillating electric field is given by
\begin{align}
& {\vec \Xi}(\tau) =\big[E_y\sin(\omega \tau),   E_x\sin(\omega \tau + \phi) , 0\big].
\label{osci}
 \end{align}
 
 The resulting time-dependent Hamiltonian is given by
\begin{align}
\mathcal H(\tau)&=J\sum_{ \la ij\ra,\ell} \Big[ S_{i,\ell}^{z}S_{j,\ell}^{z}+\frac{1}{2}\Big(S_{i,\ell}^{+}S_{j,\ell}^{-}e^{i\Phi_{ij,\ell}^s(\tau)} + {\rm H.c.}\Big)\Big] + J_c\sum_{i} {\vec S}_{i}^T\cdot{\vec S}_{i}^{B},
\end{align}
where $S^{\pm(\ell)}_{i}= S^{x}_{i,\ell} \pm i S^{y}_{i,\ell}$ denote the spin raising and lowering  operators. Note that the interlayer coupling is not affected by light intensity.   The spin current can be derived as \bea J^S = \frac {\partial\mathcal H (\tau)}{\partial \Phi_{ij,\ell} (\tau)}\equiv\sum_{j\in i; \ell}j^s_{ij,\ell},\eea where  $j^s_{ij,\ell} = -i\frac{J}{2}e^{i\Phi_{ij,\ell} (\tau)}S_{i,\ell}^{-}S_{j,\ell}^{+} + {\rm H.c.}$ Thus, the time-dependent Aharonov-Casher phase $\Phi_{ij}(\tau)$ acts as a vector potential or gauge field to the spin current.

The Floquet theory is a powerful mechanism to study periodically driven quantum systems \cite{foot3,foot4,foot5,gru,fot,fot1,jot,fla,we1,we2,we3,we4,we5,we6, gol,buk,eck1,ste,ple,ew, dik, lin, du,du1, delp, eza, zhai, saha,roy,roy1}. It enables one to  transform a time-dependent periodic Hamiltonian into a static effective Hamiltonian governed by  the Floquet Hamiltonian.  In the off-resonant limit, when the photon energy $\hbar\omega$ is greater than the energy scale of the static system, the effective static Hamiltonian is given by\cite{delp,foot4,fot1}
\bea 
\mathcal H_{eff}\approx \mathcal H_{0}+ \Delta\mathcal H_{eff},
\eea where $\Delta\mathcal H_{eff}=\big[\mathcal H_{1}, \mathcal H_{-1}\big]/\hbar\omega$  is the photon emission and absorption term.  We use the discrete Fourier component of the time-dependent Hamiltonian $\mathcal H_n=\frac{1}{T}\int_0^T d\tau e^{-in\omega \tau} \mathcal H(\tau)$ with period  $T=2\pi/\omega$. For circularly-polarized light $\phi=\pi/2$ and $E_x=E_y=E_0$, we obtain  \begin{align}
\mathcal H_n&=J\sum_{\la ij\ra,\ell}\Big[\frac{\mathcal J(\mathcal E_0)}{2}\lb S_{i,\ell}^- S_{j,\ell}^+ e^{-in\theta_{ij,\ell}}+ {\rm H.c.}\rb + \delta_{n,0}S_{i,\ell}^zS_{j,\ell}^z\Big] + J_c\delta_{n,0}\sum_{i} {\vec S}_{i}^T\cdot{\vec S}_{i}^{B},
\label{flo1}
\end{align}
where $\mathcal E_0 =  g\mu_B E_0 a/\hbar c^2$ is the dimensionless Floquet parameter,   $\theta_{ij,\ell}$ is the relative angle between ${\vec r}_i$ and  ${\vec r}_j$, $\mathcal J_n(x)$ is the Bessel function of order $n\in \mathbb Z$, and $\delta_{n,\ell}=1$ for $n=\ell$ and zero otherwise.  The zeroth-order term is given by 
\begin{align}
 \mathcal H_{0} = \sum_{ \la ij\ra,\ell}\Big[J\mathcal J_{0}(\mathcal E_0) \lb S_{i,\ell}^xS_{j,\ell}^x +S_{i,\ell}^yS_{j,\ell}^y \rb + JS_{i,\ell}^zS_{j,\ell}^z\Big] + J_c\sum_{i} {\vec S}_{i}^T\cdot{\vec S}_{i}^{B},
\end{align}
which is an $XXZ$ Heisenberg spin model, where $J\mathcal J_{0}(\mathcal E_0)< J$ for $\mathcal E_0\neq 0$. The first-order term $\Delta\mathcal H_{eff}$ involves the commutation relation $\big[ S_\alpha^+ S_\beta^-, S_\rho^+ S_{\gamma}^-\big]=2\big( \delta_{\beta\rho}S_{\beta}^zS_\alpha^+ S_\gamma^- - \delta_{\alpha\gamma}S_{\alpha}^zS_\rho^+ S_\beta^-\big)$, which gives rise to a photoinduced DM interaction \cite{owe0} of the form 
\begin{align}
\Delta\mathcal H_{eff} = D_F\sum_{\la\la ijk\ra\ra,\ell}\nu_{jk}  {\vec S}_{i,\ell}\cdot({\vec S}_{j,\ell}\times {\vec S}_{k,\ell}),
\label{DMI}
\end{align}
where $ {\vec S}_{i,\ell} = S_{i,\ell}^z\hat{z}$, $D_F=\sqrt{3}[J\mathcal J_1(\mathcal E_0)]^2/\hbar\omega$, and $\nu_{jk}=\pm 1$ for the two triangular plaquettes on the next-nearest neighbour bonds of the honeycomb lattice. Therefore, time-reversal symmetry of each honeycomb layer is broken by circularly-polarized light through a  photoinduced DM interaction, but the bilayer antiferromagnetic system still preserves $\mathcal{PT}$ symmetry. Thus, magnonic Floquet quantum spin Hall insulator can arise in irradiated bilayer collinear antiferromagnets. On the contrary, linearly-polarized light does not break time-reversal symmetry, thus $\Delta\mathcal H_{eff}=0$ for  $\phi=0$. 

 \vspace{10px}
\noindent\textbf{Periodically-driven bosonic BdG model.}~~ 
In this section, we will study the magnon band structures for a general light polarization  $\phi\in \big[0,2\pi\big]$ and a general  amplitude $E_x\neq E_y$. It is advantageous to periodically drive the bosonic BdG Hamiltonian in Eqs.~\eqref{honn} and \eqref{honn1}. In this case, the  Aharanov-Casher phase enters the momentum space Hamiltonian through the time-dependent Peierls substitution $\vec k \to \vec k - \hat{s}\mu_m {\vec \Xi}(\tau)$. Using the analysis outline in Methods, we have obtained the Fourier decomposition of the single particle bosonic BdG Hamiltonian, which enters the time-dependent Schr\"{o}dinger equation. For the  $S_z=+1$ sector, the Fourier Hamiltonian for case $(i)$ (i.e.~$J<0, J_c >0$) is given by

\begin{align}
\mathcal H_{+,q}^{(i)}(\bo)&= 
\begin{pmatrix}
|v_0|\delta_{q,0} &-\rho_{-q}^*(\vec k)& v_{J_c}\delta_{q,0}&0\\
-\rho_{q}(\vec k) &|v_0|\delta_{q,0}& 0&v_{J_c}\delta_{q,0}\\
v_{J_c}\delta_{q,0} &0& |v_0|\delta_{q,0} &-\rho_{q}(\vec k)\\
0 & v_{J_c}\delta_{q,0}&-\rho_{-q}^*(\vec k)&|v_0|\delta_{q,0}
\end{pmatrix},
\label{fhonn}
\end{align}

and  the Fourier Hamiltonian for case $(ii)$ ({\it i.e.}~$J>0, J_c >0$) is given by

\begin{align}
\mathcal H_{+,q}^{(ii)}(\bo)&= 
\begin{pmatrix}
v_0\delta_{q,0} &0& \rho_{-q}^*(\vec k)&v_{J_c}\delta_{q,0}\\
0 &v_0\delta_{q,0}& v_{J_c}\delta_{q,0}&\rho_{q}(\vec k)\\
\rho_{q}(\vec k) &v_{J_c}\delta_{q,0}& v_0\delta_{q,0}&0\\
v_{J_c}\delta_{q,0} & \rho_{-q}^*(\vec k)&0&v_0\delta_{q,0}
\end{pmatrix},
\label{fhonn1}
\end{align}
where $q\in \mathbb{Z}$ and $\rho_q(\bo)=\sum_\ell t_{\ell, q} e^{ik_\ell}$.  The renormalized Heisenberg exchange  interactions  are  given by 
\begin{align}
& t_{1,q}=|v_J|\mathcal  J_{-q}(\mathcal{E}_{-})e^{-iq\Psi_{-}},~ t_{2,q}=|v_J|\mathcal J_q(\mathcal{E}_{+})e^{iq\Psi_{+}},~ t_{3,q}= |v_J|\mathcal J_{q}(\mathcal{E}_{x})e^{iq\phi}.
\end{align}

\begin{align}
 &\mathcal{E}_{\pm}=\frac{1}{2}\sqrt{3\mathcal{E}_y^2+\mathcal{E}_x^2\pm 2\sqrt{3}\mathcal{E}_x \mathcal{E}_y\cos(\phi)} \quad{\rm and}\quad  \Psi_{\pm}=\arctan\lb \frac{\mathcal{E}_x\sin(\phi)}{\sqrt{3}\mathcal{E}_y\pm \mathcal{E}_x\cos(\phi)}\rb.
 \end{align}
In the off-resonant limit $\hbar\omega\gg J, J_c$, the system can be described by an effective time-independent  Hamiltonian given by \cite{delp,foot4,fot1}
\begin{align}
\mathcal H_{+}^{{eff}}(\vec k)&\approx \mathcal H_{+,0}(\vec k)-\frac{1}{\hbar\omega}\big[\mathcal H_{+, -1}(\vec k), \mathcal H_{+, 1}(\vec k)\big].
\label{effHam}
\end{align}
The lower block  effective  Hamiltonian is $\mathcal H_{-}^{{eff}}(\vec k) = \big[\mathcal H_{+}^{{eff}}(-\vec k)\big]^*$. The commutator term in Eq.~\eqref{effHam} contains terms proportional to $\sin(\vec k)\sin(\phi)$, which is a mass term to the Dirac magnon. This corresponds to the  momentum space of the dynamical DM interaction in Eq.~\eqref{DMI} for $\phi=\pi/2$, and it changes sign in the lower block Hamiltonian $\mathcal H_{-}^{{eff}}(\vec k)$.

In Fig.~\eqref{quasienergy}(c) and Fig.~\eqref{quasienergy} (d) we have shown the plots of the Floquet magnon quasienergies for $\phi =\pi/2$ in case $(i)$ and case $(ii)$ respectively. In both cases, we can see that the Goldstone modes are quadratically gapped out  due to SU(2) breaking anisotropy generated by radiation. In addition, the antiferromagnetic Dirac magnon at equilibrium  also becomes massive because circularly-polarized electric field induces a dynamical DM interaction in each honeycomb layer,  which preserves the $\mathcal{PT}$ symmetry of the bilayer antiferromagnetic system as shown in Eq.~\eqref{DMI}. Therefore, irradiated bilayer antiferromagnetic system  is a concrete example of a magnonic Floquet quantum spin Hall insulator, unlike irradiated 2D  graphene \cite{foot3, foot4} and 2D insulating ferromagnets \cite{owe0, owe1,owe2}.

 \vspace{10px}
 
\noindent\textbf{$\pmb{\mathbb{Z}_2}$ Topological magnon phase transition.}~~ In this section, we will study the topological phase diagram and the topological invariant quantity of the irradiated bilayer antiferromagnetic system.   Due to $S_z$ conservation, we can define the block Chern number of the Floquet magnon bands  as
 \begin{align}
 n_\sigma(\mathcal E_i,\phi) =\frac{1}{2\pi} \int d^2 k ~\Omega_\sigma(\vec k,\mathcal E_i,\phi),
 \end{align}
where  $\Omega_\sigma(\vec k,\mathcal E_i,\phi)$ are the Berry curvatures and $\sigma=\pm 1$ for $S_z=\pm $ sectors. We can then define the Hall $n_H$  and spin $n_S$ Chern numbers as
\begin{align}
&n_H(\mathcal E_i,\phi) = n_+(\mathcal E_i,\phi) + n_-(\mathcal E_i,\phi) \quad{\rm and}\quad n_S(\mathcal E_i,\phi) = \frac{1}{2}\big[ n_+(\mathcal E_i,\phi) - n_-(\mathcal E_i,\phi)\big].
\end{align}
The $\mathbb{Z}_2$ topological invariant is given by
\begin{figure}
\centering
\includegraphics[width=.9\linewidth]{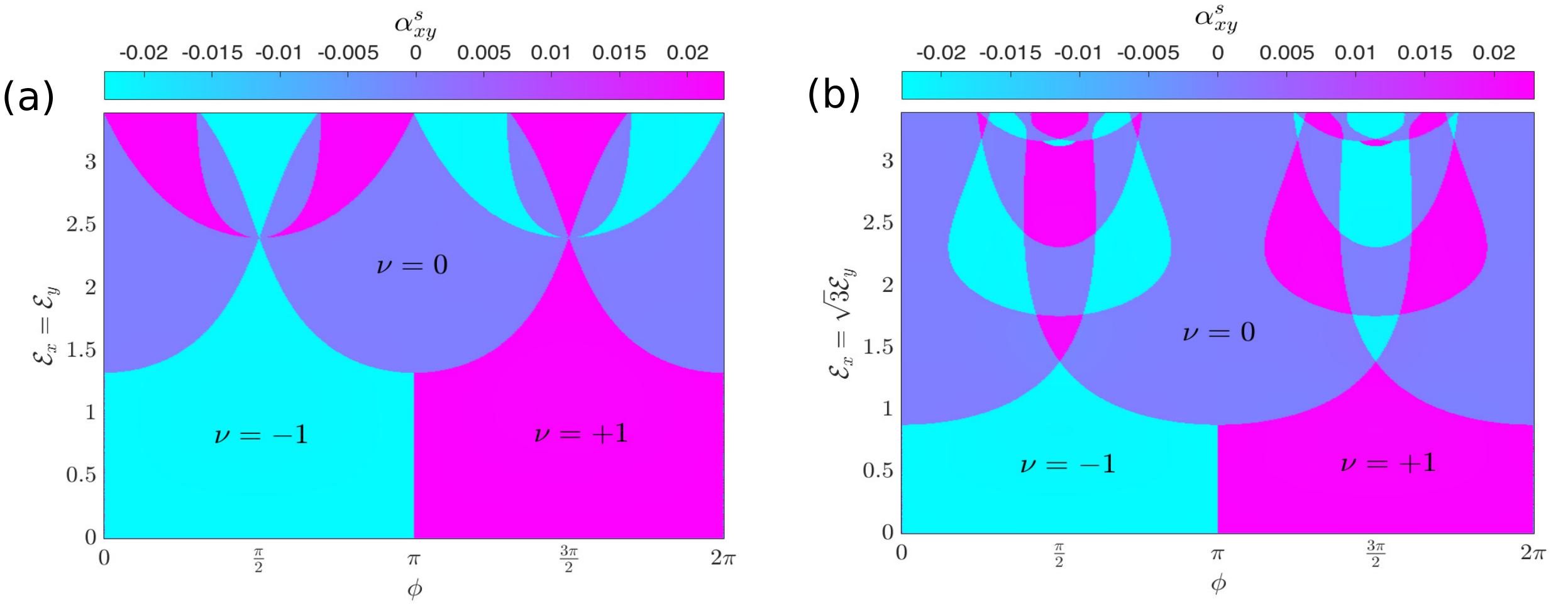}
\caption{$\mathbb{Z}_2$ Floquet topological invariant magnon phase diagram for the lower Floquet magnon band in the irradiated 2D insulating bilayer honeycomb antiferromagnets  at high frequency regime  $\hbar\omega/J = 10$ with $J_c/J=0.5$. (a) $\mathcal E_x = \mathcal E_y$. (b) $\mathcal E_x = \sqrt{3}\mathcal E_y$. The colorbar labels the Floquet magnon spin Nernst coefficient at $T/J =0.5$. Both panels correspond to  case $(ii)$. }
\label{spin_ChernN}
\end{figure}
\begin{figure}
\centering
\includegraphics[width=.9\linewidth]{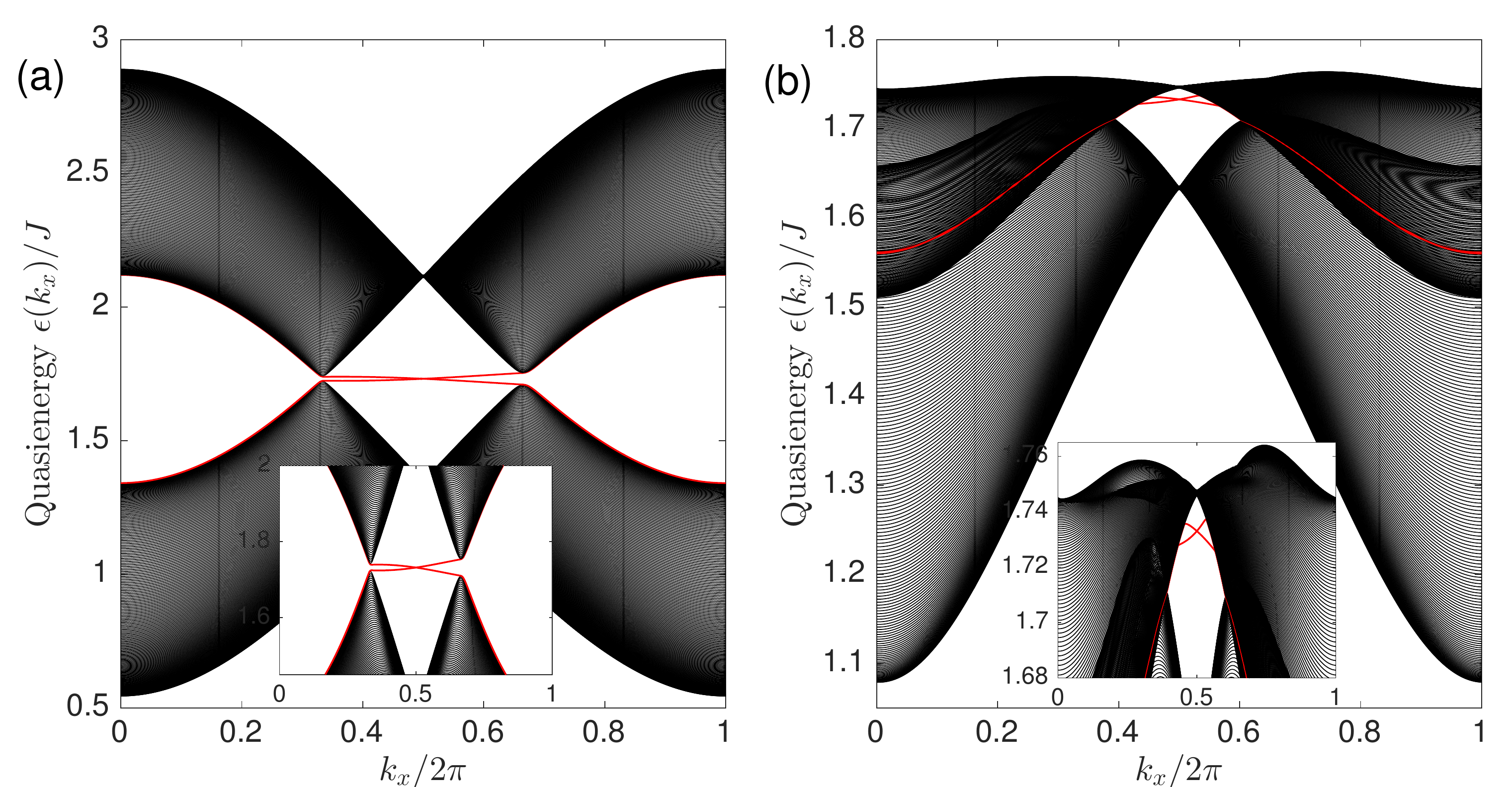}
\caption{Color online. Floquet spin filtered magnon edge states (red lines) for (a) case ($i$) and (b) case ($ii$). Insets show magnification of the Floquet spin filtered magnon edge states. The parameters are $J_c/J = 0.5$, $S=1/2$, $\mathcal{E}_x=\mathcal{E}_y = 1$ in units of $g\mu_Ba/\hbar c^2$, $\phi = \pi/2$, and $\hbar\omega/J = 10$.}
\label{edge_state}
\end{figure}

\begin{align}
\nu = n_S(\mathcal E_i,\phi)~\text{mod}~2.
\end{align}
We have computed the block Chern numbers of the system using the discretized Brillouin zone method \cite{fuk}. We  focus on the lower Floquet quasienergy magnon band. The  upper Floquet magnon band can be obtained by flipping the signs.  Due to $\mathcal{PT}$ symmetry, 
\bea
 n_+(\mathcal E_i,\phi) =- n_-(\mathcal E_i,\phi),
 \eea
hence
\begin{align}
n_H(\mathcal E_i,\phi) = 0 \quad{\rm and}\quad \nu = n_S(\mathcal E_i,\phi) =-1, 0, +1
\end{align}
for both case $(i)$ and case $(ii)$.   The vanishing of $n_H(\mathcal E_i,\phi)$ implies that the magnon thermal Hall conductivity $\kappa_{xy}^s$  \cite{alex0, alex2, alex2a, alex4h}  also vanishes in the Floquet bilayer antiferromagnetic system. However, the magnon spin Nernst conductivity \cite{z1,ran,ran1} is nonzero in the single-layer undriven system with DM interaction. Therefore, we will compute the Floquet magnon spin Nernst conductivity for the periodically driven  system.    We will focus on the regime where the  Bose occupation function is close to thermal equilibrium. Hence, we can apply the  same formula for the magnon spin Nernst coefficient  in undriven system \cite{ran1, alex2, alex2a, alex4h}, which we write as a function of the light parameters $\mathcal E_i$ and $\phi$,
\begin{align}
\alpha_{xy}^s(\mathcal E_i,\phi)= \sum_{\sigma=\pm}\int \frac{d^2 k}{(2\pi)^2}\sigma c_1\big[g(\epsilon_{\vec k, \alpha}^\sigma)\big]\Omega_\sigma^{\alpha}(\vec k,\mathcal E_i,\phi),
\end{align}
where  $c_1(x)=(1+x)\ln(1+x)-x\ln x$ is the weight function and $g(\epsilon_{\vec k, \alpha}^\sigma) = \big[ \exp{\big(\epsilon_{\vec k,\alpha}^{\sigma}/k_BT\big)} -1\big]^{-1}$  is the Bose occupation function close to thermal equilibrium and $\alpha$ labels the Floquet magnon bands. Indeed, the Floquet magnon spin Nernst coefficient is simply the $\mathbb{Z}_2$ Floquet topological invariant weighed by the $c_1(x)$ function.  Note that case $(i)$ can be regarded as two copies of the Floquet Chern insulators \cite{owe0} with opposite spins, but case $(ii)$ is not. Therefore, for case $(ii)$ we have shown in  Fig.~\eqref{spin_ChernN} the $\mathbb{Z}_2$ Floquet topological invariant magnon phase diagram of the system for the lower  Floquet quasienergy magnon band. A similar phase diagram can be obtained for case $(i)$.  We can clearly see that in the three regimes where the $\mathbb{Z}_2$ Floquet topological invariant changes sign, the Floquet spin Nernst coefficient also changes sign.

To further substantiate the existence of the $\mathbb{Z}_2$ Floquet topological invariant, we show in Fig.~\eqref{edge_state}  the plot of the Floquet magnon bands for a cylindrical strip geometry periodic along the $y$ direction and infinite along the $x$ direction.   We can see that the Floquet  magnon edge states traversing the bulk gap are spin filtered. In other words, Floquet magnon with opposite spin propagates in opposite directions.  This results in a non-vanishing Floquet magnon spin Nernst coefficient with a $\mathbb{Z}_2$ Floquet topological invariant.

\vspace{10px}
\noindent\textbf{Conclusion}

\noindent 
Using a combination of the Floquet theory, spin-wave theory, and quantum field theory, we have presented an exposition of $\mathbb{Z}_2$ magnonic Floquet quantum spin Hall states in irradiated 2D insulating bilayer honeycomb ferromagnets and antiferrmagnets which are coupled antiferromagnetically.  These systems have zero net magnetization and preserve $\mathcal{PT}$ symmetry.  In stark contrast to  irradiated graphene \cite{foot3, foot4} and insulating collinear ferromagnets \cite{owe0, owe1,owe2}, we showed that irradiation by circularly-polarized electric field breaks the time-reversal of each honeycomb layer through a photoinduced DM interaction, but  $\mathcal{PT}$ symmetry of the bilayer antiferromagnetic system is preserved.  This results in Floquet spin filtered magnon edge states protected by the $\mathbb{Z}_2$ Floquet topological invariant of the bulk magnon band. The irradiated bilayer antiferromagnetic system exhibits a non-vanishing Floquet magnon spin Nernst effect, whereas the Floquet magnon thermal Hall effect vanishes due to $\mathcal{PT}$ symmetry.  As we mentioned previously, the bilayer honeycomb chromium iodide CrI$_3$ \cite{Huang, Huang1, Huang2} is a promising candidate for investigating the current theoretical predictions. We note that  topological magnons were recently reported in the ferromagnetic bulk structure of  CrI$_3$ \cite{chen}.  Therefore, we believe that  the current predicted results are pertinent to experiments and will remarkably impact future research in topological insulating antiferromagnets and their potential practical applications to  magnon spintronics \cite{magn, benja} and  photo-magnonics \cite{benj}.

\vspace{10px}
\noindent\textbf{\large Methods} 

\noindent\textbf{Spin wave theory of bilayer honeycomb antiferromagnets.}~ To derive the bosonic Hamiltonian in Eq.~\eqref{bosonicH}, we introduce the Holstein Primakoff  bosons:  
\begin{align}
&S_{i}^{z(\ell)}= S-a_{i}^{(\ell)\dagger} a_{i}^{(\ell)},~S_{i}^{+(\ell)}= \sqrt{2S}a_{i}^{(\ell)}=(S_{i}^{-(\ell)})^\dg,
\end{align}
for up pointing spins, and
\begin{align}
&S_{i}^{z(\ell)}= -S+a_{i}^{(\ell)\dagger} a_{i}^{(\ell)},~S_{i}^{+(\ell)}= \sqrt{2S}a_{i}^{\dg(\ell)}=(S_{i}^{-(\ell)})^\dg,
  \end{align}
  for down pointing spins. 
  
Here $a_{i}^{(\ell)\dagger} (a_{i}^{(\ell)})$ are the bosonic creation (annihilation) operators on the sublattices and  $S^{\pm(\ell)}_{j}= S^{x(\ell)}_{j} \pm i S^{y(\ell)}_{j}$ denote the spin raising and lowering  operators. The bosonic tight-binding models are given by  
  
  \begin{align}
  \mathcal H &= JS\sum_{\la ij\ra, \ell }\big[\big(a_{i}^{(\ell)}a_{i}^{(\ell)\dg} + a_{j}^{(\ell)}a_{j}^{(\ell)\dg}\big)-\big(a_{i}^{(\ell)} a_{j}^{(\ell)\dg}+\rm{H.c.}\big)\big]\nonumber\\& +\frac{J_cS}{2}\sum_{i, \ell\neq \ell^\prime}\big[\big(a_{i}^{(\ell)}a_{i}^{(\ell)\dg} + a_{i}^{(\ell^\prime\big)}a_{i}^{(\ell^\prime)\dg})+\big(a_{i}^{(\ell)\dg} a_{i}^{(\ell^\prime)\dg}+\rm{H.c.}\big)\big],
  \label{eq25}
  \end{align}
for case $(i)$ and

  \begin{align}
  \mathcal H &= JS\sum_{\la ij\ra, \ell }\big[\big(a_{i}^{(\ell)}a_{i}^{(\ell)\dg} + a_{j}^{(\ell)}a_{j}^{(\ell)\dg}\big)+\big(a_{i}^{(\ell)\dg} a_{j}^{(\ell)\dg}+\rm{H.c.}\big)\big]\nonumber\\& +\frac{J_cS}{2}\sum_{i, \ell\neq \ell^\prime}\big[\big(a_{i}^{(\ell)}a_{i}^{(\ell)\dg} + a_{i}^{(\ell^\prime)}a_{i}^{(\ell^\prime)\dg}\big)+\big(a_{i}^{(\ell)\dg} a_{i}^{(\ell^\prime)\dg}+\rm{H.c.}\big)\big],
  \label{eq26}
  \end{align}
for case $(ii)$. The Fourier transform of Eqs.~\eqref{eq25} and \eqref{eq26} gives the bosonic BdG Hamiltonian in Eq.~\eqref{bosonicH}.

\vspace{10px}
\noindent\textbf{Quantum field theory description of Aharonov-Casher phase.}~ In the presence of an electromagnetic field,  the low-energy charge-neutral Dirac magnon near the ${\bf K}$-point of the BZ  is governed  by the 2+1 dimensional Dirac-Pauli Lagrangian  \cite{hagen, bjo}  

\begin{align}
 \mathcal L=\bar\psi\Big[-\varepsilon_0\gamma^0 + v_Di\gamma^\mu\partial_\mu-\frac{v_D\mu_m}{2}\sigma^{\mu\nu} F_{\mu\nu}\Big]\psi,
 \label{Lag}
\end{align}
where $\varepsilon_0$ accounts for the finite energy Dirac magnon and $v_D$ is the group velocity of the Dirac magnon, and $\bar\psi=\psi^\dg\gamma^0$.  The electromagnetic field tensor is $F_{\mu\nu}$   and  $\sigma^{\mu\nu}=\frac{i}{2}[\gamma^\mu,\gamma^\nu]=i\gamma^\mu\gamma^\nu,~ (\mu\neq \nu)$ with $\gamma^{\mu}=(\gamma^0,\gamma^i)$. 

 To describe the  antiferromagnetic Dirac magnon in the presence of an oscillating electric field, we follow the procedure in ref.~\cite{xia}. In (2+1) dimensions, there are two inequivalent representations of the Dirac gamma matrices which generate different Clifford algebras. These two inequivalent representations of the Dirac matrices can be used to describe the $S_z =1$ and $S_z =-1$ sectors of the antiferromagnetic Dirac magnons. They obey the relation 
 \begin{align}
 \gamma^\mu\gamma^\nu  = g^{\mu\nu} + i\hat{s}\epsilon^{\mu\nu\lambda}\gamma_\lambda
 \end{align}
where $g^{\mu\nu}=\text{diag}(1,-1,-1)$ is the Minkowski metric  and $\epsilon^{\mu\nu\lambda}$ is an antisymmetric tensor in (2+1) dimensions. 
\begin{align}
\hat s = i\gamma^0\gamma^1\gamma^2 = \gamma^0\sigma^{12},
\end{align}
with eigenvalues $s = \pm 1$. We choose the representation
\begin{align}
&\gamma^0 = 
\begin{pmatrix}
 \sigma_z& 0\\
 0&\sigma_z
 \end{pmatrix},
 \quad 
 \gamma^1 = 
\begin{pmatrix}
 i\sigma_y& 0\\
 0&-i\sigma_y
 \end{pmatrix},
 \quad 
 \gamma^2 = 
\begin{pmatrix}
 -i\sigma_x& 0\\
 0&-i\sigma_x
 \end{pmatrix},
\end{align}
such that $\hat s$ is diagonal and it is given by

\begin{align} 
 \hat s = 
\begin{pmatrix}
 I& 0\\
 0&-I
 \end{pmatrix}.
 \end{align}  
In this representation the interaction term transforms as
\begin{align}
\bar\psi \sigma^{\mu\nu} F_{\mu\nu}\psi = -\hat s\epsilon_{\mu\nu\lambda}F^{\mu\nu}\bar\psi\gamma^\lambda\psi.
\end{align}
The Lagrangian can then be written as
\begin{align}
 \mathcal L=\bar\psi\Big[-\varepsilon_0\gamma^0 + iv_D\gamma^\mu\partial_\mu + \hat sv_D\mu_m\gamma^{\mu} Q_{\mu}\Big]\psi,
 \label{Lag1}
\end{align}
where $Q_{\mu} = (1/2)\epsilon_{\lambda\nu\mu}F^{\lambda\nu}$ is the effective vector potential dual of the field strength tensor. We consider an electromagnetic field with only an oscillating  electric field vector $\vec{E}(\tau)$. Hence, $Q_{\mu} = \Xi(\tau) = \vec{E}(\tau)\times z$. The Hamiltonian is given by
 \begin{align}
 H = \int d^2 x ~\big [ \pi(x)\dot{\psi}(x) - \mathcal L \big ] \equiv \int d^2 x ~ \psi^\dg \mathcal H_D \psi,
 \end{align}
 where $\pi(x) = \frac{\partial \mathcal L}{\partial{\dot{\Psi}(x)}}$ 
is the generalized momentum. The  Hamiltonian is given by
\begin{align}
 \mathcal H_D=v_0 + v_D \vec{\alpha}\cdot \lb -i\vec{\nabla}- \hat s\mu_m\Xi(\tau) \rb,
 \label{hamil}
\end{align}
where $\vec{\alpha}=\gamma^0\vec{\gamma}$. This is the effective form of the bosonic BdG Hamiltonian in Eq.~\eqref{bosonicH} in the presence of an oscillating  electric field vector.

\vspace{10px}
\noindent\textbf{Floquet-Bloch theory.}~~  The time-dependent bosonic BdG Hamiltonian $\mathcal H(\vec{k},\tau)$ can be studied by the Floquet-Bloch formalism. We can expand it  as
\begin{align}
\mathcal H(\vec{k},\tau)=\sum_{n=-\infty}^{\infty} e^{in\omega \tau}\mathcal H_{n}(\vec{k}),
\end{align}
where the Fourier components  are given by $\mathcal H_{n}(\vec{k})=\frac{1}{T}\int_{0}^T e^{-in\omega \tau}\mathcal H(\vec{k}, \tau) d\tau=\mathcal H_{-n}^\dg(\vec{k}).$ The corresponding  eigenvectors can be written as $\ket{\psi_\alpha(\vec{k}, \tau)}=e^{-i \epsilon_\alpha(\vec{k}) \tau}\ket{u_\alpha(\vec{k}, \tau)},$
where  $\ket{u_\alpha(\vec{k}, \tau)}=\ket{u_\alpha(\vec{k}, \tau+T)}=\sum_{n} e^{in\omega \tau}\ket{u_{\alpha}^n(\vec{k})}$ is the time-periodic Floquet-Bloch wave function of magnons and $\epsilon_\alpha(\vec{k})$ are the magnon quasi-energies. We define the Floquet operator  as $\mathcal H^F(\vec{k},\tau)=\mathcal H(\vec{k},\tau)-i\partial_\tau$.  The corresponding eigenvalue equation is of the form 
\begin{align}
\sum_m \big[\mathcal H_{n-m}(\vec{ k}) + m\omega\delta_{n,m}\big]u_{\alpha}^m(\vec{k})= \epsilon_\alpha(\vec{k})u_{\alpha}^n(\vec{k}).
\end{align}
Each block Hamiltonian in Eq.~\eqref{bosonicH} obeys this equation.

\vspace{10px}
\noindent\textbf{Acknowledgements}

\noindent Research at Perimeter Institute is supported by the Government of Canada through Industry Canada and by the Province of Ontario through the Ministry of Research
and Innovation. 

\vspace{10px}
\noindent\textbf{Author Contributions}

\noindent S. A. Owerre conceived the idea, performed the calculations, discussed the results, and wrote the manuscript.

\vspace{10px}
\noindent\textbf{Additional Information}

\noindent\textbf{\small Competing Interests}. I declare that the author has no competing interests as defined by Nature Research, or other interests that might be perceived to influence the results and/or discussion reported in this paper.


\vspace{10px}
\textbf{References}
 \vspace{-40px}
 
\begin{thebibliography}{99}

\bibitem{foot3}
 Oka, T. \&    Aoki, H. Photovoltaic Hall effect in graphene. Phys. Rev. B {\bf 79}, 081406 (2009).

\bibitem{foot5}
 Inoue, J. -I. \&  Tanaka, A. Photoinduced Transition between Conventional and Topological Insulators in Two-Dimensional Electronic Systems. Phys. Rev. Lett. {\bf 105}, 017401 (2010).
 
\bibitem{lin}
Lindner, N.,  Refael, G. \&  Gaslitski, V. Floquet topological insulator in semiconductor quantum wells. Nat. Phys. {\bf 7}, 490 (2011).

\bibitem{fot}
 Calvo H. L. et al. Tuning laser-induced bandgaps in graphene.  Appl. Phys. Lett. {\bf 98}, 232103 (2011).
\bibitem{foot4}
 Kitagawa T. et al. Transport properties of nonequilibrium systems under the application of light: Photoinduced quantum Hall insulators without Landau levels.  Phys. Rev. B {\bf 84}, 235108 (2011).
\bibitem{delp}
 Delplace, P.,    G\'omez-Le\'on, \'A. \& Platero, G. Merging of Dirac points and Floquet topological transitions in ac-driven graphene.  Phys. Rev. B {\bf 88}, 245422 (2013).
\bibitem{fot1}
 Cayssol J.  et al. Floquet topological insulators.  Physica Status Solidi (RRL) {\bf 7}, 101 (2013).
\bibitem{dik}
 Wang Y. H. et al. Observation of Floquet-Bloch States on the Surface of a Topological Insulator. Science  {\bf 342}, 453 (2013)

\bibitem{ew}
 Rechtsman, M. C. Photonic Floquet topological insulators.  Nature {\bf 496}, 196 (2013).
\bibitem{eza}
 Ezawa, M. Photoinduced Topological Phase Transition and a Single Dirac-Cone State in Silicene. Phys. Rev. Lett. {\bf 110}, 026603 (2013).

\bibitem{gru}
 Grushin, A. G.,   G\'omez-Le\'on, \'A. \&  Neupert, T. Floquet Fractional Chern Insulators. Phys. Rev. Lett. {\bf 112}, 156801 (2014).
 
 \bibitem{jot}
 Jotzu,  G. et al. Experimental realization of the topological Haldane model with ultracold fermions.  Nature {\bf 515}, 237 (2014).
  \bibitem{zhai}
 Zhai,  X. \&  Jin, G. Photoinduced topological phase transition in epitaxial graphene. Phys. Rev. B {\bf 89}, 235416 (2014).
  \bibitem{fla}
Fl\"aschner,  N.  et al. Experimental reconstruction of the Berry curvature in a Floquet Bloch band. Science {\bf 352}, 1091 (2016).
   
  \bibitem{we1}
 Wang R. et al.  Floquet Weyl semimetal induced by off-resonant light. EPL (Europhys. Lett.) {\bf 105}, 17004 (2014). 
\bibitem{gol}
 Goldman,  N. \&  Dalibard, J. Periodically Driven Quantum Systems: Effective Hamiltonians and Engineered Gauge Fields. Phys. Rev. X {\bf 4}, 031027 (2014).
   \bibitem{buk}
 Bukov,  M.,  D'Alessio, L. \&  Polkovnikov, A. Universal High-Frequency Behavior of Periodically Driven Systems: from Dynamical Stabilization to Floquet Engineering.  Adv. Phys.  {\bf 64}, 139,  (2015).
  \bibitem{eck1}
 Eckardt,  A. \&  Anisimovas, E. High-frequency approximation for periodically driven quantum systems from a Floquet-space perspective. New J. Phys. {\bf 17},  093039 (2015).
 \bibitem{we2}
 Ebihara, S.,  Fukushima, K. \&  Oka, T. Chiral pumping effect induced by rotating electric fields. Phys. Rev. B {\bf 93}, 155107 (2016).
 \bibitem{we3}
 Chan C. -K. et al. When Chiral Photons Meet Chiral Fermions: Photoinduced Anomalous Hall Effects in Weyl Semimetals.  Phys. Rev. Lett. {\bf 116}, 026805 (2016).
\bibitem{we4}
 Yan,  Z. \&  Wang, Z. Tunable Weyl Points in Periodically Driven Nodal Line Semimetals. Phys. Rev. Lett. {\bf 117}, 087402 (2016).
\bibitem{we5}
 Zhang, X. -X.,  Ong, T. T. \&  Nagaosa, N. Theory of photoinduced Floquet Weyl semimetal phases. Phys. Rev. B {\bf 94}, 235137 (2016).
\bibitem{saha}
 Saha, K. Photoinduced Chern insulating states in semi-Dirac materials. Phys. Rev. B {\bf 94}, 081103(R) (2016).
\bibitem{we6}
 H\"ubener, H. et al. Creating stable Floquet-Weyl semimetals by laser-driving of 3D Dirac materials.  Nat. Commun. {\bf 8}, 13940 (2017).

  \bibitem{ste}
 Stepanov,  E. A.,  Dutreix, C. \&   Katsnelson, M. I. Dynamical and Reversible Control of Topological Spin Textures. Phys. Rev. Lett. {\bf 118}, 157201 (2017).


 \bibitem{ple}
 Plekhanov,  K.,  Roux, G. \& Le Hur,  K. Floquet engineering of Haldane Chern insulators and chiral bosonic phase transitions. Phys. Rev. B {\bf 95}, 045102 (2017).

 \bibitem{du}
 Du,  L.,  Zhou,  X. \&  Fiete, G. A. Quadratic band touching points and flat bands in two-dimensional topological Floquet systems. Phys. Rev. B {\bf 95}, 035136 (2017).
 \bibitem{du1}
 Wang,  Y.,   Liu, Y. \&  Wang, B.  Effects of light on quantum phases and topological properties of two-dimensional Metal-organic frameworks. Sci. Rep. {\bf 7}, 41644  (2017).
 \bibitem{roy}
 Roy, R. \&  Harper, F. Periodic table for Floquet topological insulators. Phys. Rev. B {\bf 96}, 155118 (2017).
 \bibitem{roy1}
Yao, S.,  Yan, Z. \& Wang, Z. Topological invariants of Floquet systems: General formulation, special properties, and Floquet topological defects. Phys. Rev. B {\bf 96}, 195303 (2017).

\bibitem{magn}
  Chumak,   A. V. et al. Magnon spintronics.  Nat. Phys. {\bf 11}, 453 (2015).
  
\bibitem{benja}
 Lenk,  B. et al. The building blocks of magnonics. Phys. Rep. {\bf 507}, 107 (2011). 

  \bibitem{aha}
 Aharonov,  Y. \&  Casher, A. Topological Quantum Effects for Neutral Particles. Phys. Rev. Lett. {\bf 53}, 319 (1984).

 \bibitem{ahaz}
 Cao,  Z., Yu,  X. \&  Han, R. Quantum phase and persistent magnetic moment current and Aharonov-Casher effect in a $s=1/2$ mesoscopic ferromagnetic ring. Phys. Rev. B. {\bf 56}, 5077 (1997).
  \bibitem{mei1}
 Meier F. \&  Loss, D. Magnetization Transport and Quantized Spin Conductance. Phys. Rev. Lett. {\bf 90}, 167204 (2003).
  \bibitem{ahat}
Liu, T.  \& Vignale,  G. Electric Control of Spin Currents and Spin-Wave Logic. Phys. Rev. Lett. {\bf 106}, 247203 (2011).
  \bibitem{tak4}
 Zhang,  X.  et al. Electric-field coupling to spin waves in a centrosymmetric ferrite.  Phys. Rev. Lett. {\bf 113}, 037202 (2014).
\bibitem{mei}
 Nakata,  K.,  Klinovaja, J. \&  Loss, D. Magnonic quantum Hall effect and Wiedemann-Franz law.  Phys. Rev. B. {\bf 95}, 125429 (2017).
 
  \bibitem{xr}
 Ying  S. \&  Wang, X. R. Chiral anomaly of Weyl magnons in stacked honeycomb ferromagnets. Phys. Rev. B {\bf 96}, 104437 (2017).
   \bibitem{xr1}
  Ying  S., Wang, X. S, \&  Wang, X. R. Magnonic Weyl semimetal and chiral anomaly in pyrochlore ferromagnets. Phys. Rev. B {\bf 95}, 224403 (2017).



 


\bibitem{owe0}
 Owerre, S. A. Floquet Topological Magnons.  J. Phys. Commun. {\bf 1}, 021002 (2017).

\bibitem{owe2}
Owerre, S. A.  Photoinduced Topological Phase Transitions in Topological Magnon
Insulators. Sci. Rep. {\bf 8}, 4431 (2018).

\bibitem{kar}
 Kar, S. \&  Basu, B. Photo-induced Entanglement in a Magnonic Floquet Topological Insulator.  Phys.Rev. B {\bf 98} 245119 (2018).

\bibitem{owe1}
 Owerre, S. A. Floquet Weyl Magnons in Three-Dimensional Quantum Magnets. Sci. Rep. {\bf 8}, 10098 (2018).

  \bibitem{ment}
 Mentink, et al. Manipulating magnetism by ultrafast control of the exchange interaction. J. Phys.: Condens. Matter {\bf 29} 453001 (2017).

  \bibitem{tak4a}
 Schellekens,  A. J.  et al. Ultrafast spin-transfer torque driven by femtosecond pulsed-laser excitation.  Nat. Commun. {\bf 5}, 4333 (2014).
 \bibitem{walo}
 Walowski, J.  \& M\"{u}nzenberg,  M., Perspective: Ultrafast magnetism and THz spintronics. J. Appl. Phys. {\bf 120}, 140901 (2016).  
 
\bibitem{sol}
  Owerre, S. A. A first theoretical realization of honeycomb topological magnon insulator. J. Phys.: Condens. Matter {\bf 28}, 386001 (2016). 
 
 \bibitem{dm}
 Dzyaloshinsky,  I. A thermodynamic theory of ``weak'' ferromagnetism of antiferromagnetics. J. Phys. Chem. Solids {\bf 4}, 241 (1958).
  \bibitem{dm2}
 Moriya,    T. Anisotropic Superexchange Interaction and Weak Ferromagnetism. Phys. Rev. {\bf 120}, 91 (1960).



\bibitem{z1}
Kovalev,  A.  A. \&   Zyuzin, V.  Magnon spin Nernst effect in antiferromagnets. Phys. Rev. B {\bf 93}, 161106(R) (2016).

\bibitem{z2}
 Nakata, K. Kim, S. E, Klinovaja, J. \& Loss, D. Magnonic topological insulators in antiferromagnets. Phys. Rev. B {\bf 96}, 224414 (2017).
 
\bibitem{z3}
 Lee, K. H.,  Chung, S. B, Park, K. \&  Park, J. -G. Magnonic quantum spin Hall state in the zigzag and stripe phases of the antiferromagnetic honeycomb lattice. Phys. Rev. B 97, 180401(R) (2018).
 
 \bibitem{z4}
 Mook, A.,   G\"{o}bel, B.,  Henk, J. \& Mertig, I. Taking an electron-magnon duality shortcut from electron to magnon transport. Phys. Rev. B 97, 140401(R) (2018).
 
 \bibitem{z5}
 Kondo, H.  Akagi, Y. \& Katsura, H. $\mathbb{Z}_2$ Topological Invariant for Magnon Spin Hall Systems. Phys. Rev. B {\bf 99}, 041110 (2019).
 
  \bibitem{z6}
 Joshi, D. G \&  Schnyder, A. P.  $\mathbb{Z}_{2}$  topological quantum paramagnet on a honeycomb bilayer. 	arXiv:1809.06387 (2018).
 
 \bibitem{qs}
 Kane, C. L. \&  Mele, E. J. Quantum Spin Hall Effect in Graphene.  Phys. Rev. Lett. {\bf 95}, 226801 (2005).
 
 \bibitem{qs1}
 Kane, C. L. \&  Mele, E. J. $\mathbb{Z}_2$ Topological Order and the Quantum Spin Hall Effect.  Phys. Rev. Lett. {\bf 95}, 146802 (2005).
 
 
 \bibitem{qs2}
 Bernevig, B. A.,   Hughes, T. L. \&  Zhang, S. C. Quantum Spin Hall Effect and Topological Phase Transition in HgTe Quantum Wells. Science, {\bf 314}, 1757 (2006).

 \bibitem{shi}
 Shiomi, Y.  Takashima, R. \& E. Saitoh, E. Signature of magnon Nernst effect in an antiferromagnetic insulator. Phys. Rev. B {\bf 96}, 134425 (2017).
 
 \bibitem{Huang}
 Huang, B. et al. Layer-dependent ferromagnetism in a van der Waals crystal down to the monolayer limit. Nature (London) {\bf 546}, 270 (2017).
 
 \bibitem{Huang1}
 Jiang,  S., Shan, J. \&  Mak, K. F. Electric-field switching of two-dimensional van der Waals magnets. Nat. Mater. {\bf 17}, 406 (2018).
 
 \bibitem{Huang2}
 Klein, D. R. et  al. Probing magnetism in 2D van der Waals crystalline insulators via electron tunneling. Science {\bf 360}, 1218 (2018).
 
 \bibitem{benj}
 Lenk,  B. et al. Photo-magnonics.  arXiv:1208.5383 (2012).
 
 
 \bibitem{hp}
 Holstein, T. \&  Primakoff, H. Field Dependence of the Intrinsic Domain Magnetization of a Ferromagnet.  Phys. Rev. {\bf 58}, 1098 (1940).

 
 \bibitem{fuk}
 Fukui, T., Hatsugai,  Y. \&  Suzuki, H.  Chern Numbers in Discretized Brillouin Zone: Efficient Method of Computing (Spin) Hall Conductances. J. Phys. Soc. Jpn. {\bf 74}, 1674 (2005).
 
 \bibitem{alex0}
 Katsura,  H.,  Nagaosa, N. \&  Lee, P. A. Theory of the Thermal Hall Effect in Quantum Magnets.  Phys. Rev. Lett.  {\bf 104},  066403 (2010).
 
 
 \bibitem{alex2}
 Matsumoto,  R. \&  Murakami, S.  Theoretical Prediction of a Rotating Magnon Wave Packet in Ferromagnets. Phys. Rev. Lett. {\bf 106}, 197202 (2011).
 
   \bibitem{alex2a}
  Matsumoto,  R. \&  Murakami, S. Rotational motion of magnons and the thermal Hall effect. Phys. Rev. B {\bf 84}, 184406 (2011).
\bibitem{alex4h} 
 Lee,  H.,  Han, J. H. \&  Lee, P. A.  Thermal Hall effect of spins in a paramagnet. Phys. Rev. B.  {\bf 91},  125413 (2015).
 
 
  \bibitem{ran}
 Cheng, R.,  Okamoto, S. \&  Xiao, D. Spin Nernst Effect of Magnons in Collinear Antiferromagnets. Phys. Rev. Lett. {\bf 117}, 217202 (2016).

 \bibitem{ran1}
  Zhang, Y.,  Okamoto, S. \&  Xiao, D. Spin-Nernst effect in the paramagnetic regime of an antiferromagnetic insulator. Phys. Rev. B {\bf 98}, 035424 (2018).

\bibitem{hagen}
Hagen, C. R. Exact equivalence of spin-$1/2$ Aharonov-Bohm and Aharonov-Casher effects. Phys. Rev. Lett. {\bf 64}, 2347 (1990).
\bibitem{bjo} 
 Bjorken, J. D.  \&  Drell, S. D. \textit{Relativistic Quantum Mechanics}  (New York, McGraw-Hill) (1964).
 
 
 \bibitem{xia} 
 He., X. -G. \&  McKellar, B. H. J. Topological effects, dipole moments, and the dual current in 2+1 dimensions. Phys. Rev. A {\bf 64}, 022102 (2001).
 
 \bibitem{chen}
 Chen, L. et al. Topological spin excitations in honeycomb ferromagnet CrI$ _3$.  Phys. Rev. X {\bf 8}, 041028 (2018). 
 
\end{thebibliography}
\end{document}